\documentclass[3p,10pt,a4paper,twoside,fleqn,sort&compress]{elsarticle}
\makeatletter
\def\ps@pprintTitle{%
  \let\@oddhead\@empty
  \let\@evenhead\@empty
  \let\@oddfoot\@empty
  \let\@evenfoot\@oddfoot
} \makeatother

\usepackage{graphicx}
\usepackage{verbatim}
\usepackage{amsmath}
\usepackage[varg]{pxfonts}
\usepackage{eepic}
\usepackage{amssymb}

\usepackage{mathtools}
\usepackage[all,cmtip]{xy}

\usepackage{tikz}
\usetikzlibrary{matrix,arrows}

\newtheorem{theorem}{Theorem}[section]

\newcommand{\mc}{\mathcal}

\newcommand{\mcA}{A}

\newcommand{\mcP}{P}
\newcommand{\mcU}{U}
\newcommand{\mcQ}{Q}
\newcommand{\mcPQ}{PQ}

\newcommand{\Hs}{\mc H}
\newcommand{\Hv}{\mc H}
\newcommand{\mcT}{\mc T}

\newcommand{\isoXtoR}{h^{-1}}
\newcommand{\isoRtoX}{h}

\begin{document}

\begin{frontmatter}

\title{\Large\bf An axiomatic characterization of generalized entropies\\ under analyticity condition}
\tnotetext[t1]{Research supported by Ministry  of Science and
Technological Development, Republic of Serbia, Grants No. 174013
and 174026}

\author[misanu]{Velimir M. Ili\'c\corref{cor}}
\ead{velimir.ilic@gmail.com}

\author[fosun]{Miomir S. Stankovi\'c}
\ead{miomir.stankovic@gmail.com}

\cortext[cor]{Corresponding author. Tel.: +38118224492; fax:
+38118533014.}
\address[misanu]{Mathematical Institute of the Serbian Academy of Sciences and Arts, Kneza Mihaila 36, 11000 Beograd, Serbia}
\address[fosun]{University of Ni\v s, Faculty of Occupational Safety, \v Carnojevi\'ca 10a, 18000 Ni\v s, Serbia}

\begin{abstract}
\begin{minipage}{15.5cm}
We present the characterization of the Nath, R\'enyi and
Havrda-Charv\'at-Tsallis entropies under the assumption that they
are analytic function with respect to the distribution dimension,
unlike the the previous characterizations, which supposes that
they are expandable maximized for uniform distribution.
\end{minipage}
\end{abstract}

\end{frontmatter}

\newcommand{\Logp}{\hbox{Log}_q}
\newcommand{\Logq}{\hbox{Log}_q}
\newcommand{\Expp}{\hbox{Exp}_\op}
\newcommand{\op}{\oplus}
\newcommand{\cop}{\tau}
\newcommand{\Mp}{\hbox{M}_\op}
\newcommand{\Jp}{\mc J}

\newcommand{\ra}{\rightarrow}

\newcommand{\sR}{\mathbb R}
\newcommand{\sN}{\mathbb N}
\newcommand{\sC}{\mathbb C}
\newcommand{\sX}{\mathbb X}
\newcommand{\pvec}{(p_1, \dots, p_n)}
\newcommand{\qvec}{(q_1, \dots, q_m)}
\newcommand{\escP}{\mcP^{(\alpha)}}
\newcommand{\escpvec}{( \escp{1}, \dots, \escp{n})}
\newcommand{\escp}[1]{p_{#1}^{(\alpha)}}
\newcommand{\escr}[1]{r_{#1}^{(\alpha)}}
\newcommand{\escq}[1]{q_{#1}^{(\alpha)}}
\newcommand{\escqc}[2]{q_{#1 | #2}^{(\alpha)}}

\section{Introduction}

Nath \cite{Nath_68b}, R\'enyi \cite{Renyi_07} 
and Havrda-Charv\'at-Tsallis \cite{Havrda_Charvat_67},
\cite{Tsallis_88} entropies are well known generalizations of the
Shannon entropy. All of them have more general strong additivity
property in comparation to the Shannon entropy. By the strong
additivity property, the entropy of joint distribution can be
represented as the sum of the the entropy of the first one and the
conditional entropy of the second one with respect to the first
one. The conditional entropy is defined as the $\mcP$ expected
value of the entropy of the conditional distribution $\mcQ$
conditioned on $\mcP$. For Nath entropy, and its normalized
instance, the R\'enyi entropy, the definition of the expectation
is generalized from linear to the quasi-linear mean, while in the
Havrda-Charv\'at-Tsallis case linear expectation is used but the
additivity is generalized to the $\gamma$-additivity
\cite{Nivanen_Wang__03}.

Previous axiomatic systems \cite{Khinchin_57},
\cite{Jizba_Arimitsu_04}, \cite{Ilic_et_GSK}, \cite{Abe_00} take
additivity condition as an axiom, and another three axioms -
continuity, expandability (which means that adding the zero
probability event in the sample space does not affect the entropy
distribution) and maximality (which means that the entropy is
maximized for uniform distribution). In this letter we give
alternative axiomatic systems, which replace the expendability and
maximality axioms with the axiom that states the uniform
distribution entropy can be analytical continued if it is taken as
the function of the distribution dimension, the property that has
an important role in asymptotic analysis of entropy
\cite{Jacquet_Szpankowski_99} (see \cite{Knessl_98} for
alternative approach). The presented results generalizes
discussion in \cite{Nambiar_92}, where the Shannon entropy is
considered.

The letter is organized as follows. In section \ref{sec: Shannon}
we review the Shannon entropy uniqueness theorem given by Nambiar
et. al \cite{Nambiar_92}. The theorem is generalized to the Nath
and R\'enyi entropies in section \ref{sec: Nath} and to the
Havrda-Charv\'at-Tsallis entropy in section \ref{sec: Tsallis}.

\section{Shannon entropy}
\label{sec: Shannon}

Let the set of all $n$-dimensional distributions be denoted with
\begin{equation}
   \Delta_n \equiv \left\{ (p_1, \dots , p_n) \Big\vert \; p_i \ge 0,
      \sum_{i=1}^{n} p_i = 1 \right \}
  \label{Delta},\quad n>1
\end{equation}
and the $n$-dimensional uniform distribution be denoted with
\begin{equation}
\mcU_n=\left(\frac{1}{n}, \dots, \frac{1}{n}\right) \in \Delta_n.
\end{equation}

Shannon entropy \cite{Shannon_48} of $n$-dimensional distribution
is a function $\Hs_n: \Delta_n \ra \sR_{>0}$ given with the family
parameterized by $\cop \in \sR$:
\begin{equation}
\label{Shannon: H final c} \mc S_n(\mcP) = \cop \cdot
\sum_{k=1}^{n} p_k  \log p_k ; \quad\quad\quad \ \cop <0,
\end{equation}
where $\log$ stands for the logarithm to the base $2$.
The following theorem characterizes the Shannon entropy.

\begin{theorem}\rm
\label{Shannon: theorem}

Let the function $\Hs_n: \Delta_n \ra \sR_{>0}$ satisfies the
following axioms, for all $n \in \sN$, $n>1$:

\begin{description}

\item[SA1:] 
$\Hs_n(\mcU_n)=s(1/n)$, where $s:\sC\ra\sC$ is an analytic
function.
\item[SA2:]

Let $\mcP = (p_1, \dots, p_n) \in \Delta_n$, 
$\mcPQ = (r_{11}, r_{12}, \dots, r_{nm}) \in \Delta_{nm}$, $n, m
\in \sN$ such that $p_i = \sum_{j=1}^n r_{i j}$, and $\mcQ_{ | k}
= (q_{1|k}, \dots, q_{m|k}) \in \Delta_m$, where $q_{i|k} =
r_{ik}/p_k$. Then,
\begin{equation}
\Hs_{nm}(\mcPQ) = \Hs_n(\mcP) + \Hs_m(\mcQ| \mcP),
\quad\text{where}\quad
\Hs_m(\mcQ| \mcP) = \sum_k p_k \Hs_m(\mcQ_{ | k}))
\end{equation}

\end{description}

Then, $\Hs_n$ is the Shannon entropy.
\end{theorem}

The previous theorem slightly differs from the one presented in
\cite{Nambiar_92}, but it can be proven by straightforward
repetition of the steps from \cite{Nambiar_92}. First, we do not
assume the normalization condition $s(1/2)=1$, which fixes the
constant to $\cop=-1$. Second, the statement of the theorem from
\cite{Nambiar_92} assumes that the entropy is complex analytic
function with respect to the distribution. We assume the
equivalent statement that the entropy is continuous real function
(note that the assumption about the analyticity with respect to
the distribution dimension is kept).


\section{Nath entropy}
\label{sec: Nath}

For a distribution $\mcP \in \Delta_n$ we define the
$\alpha$-escort distribution
$\mcP^{(\alpha)}=(\escp{1},\dots,\escp{n})$, where
\begin{equation}
\escp{k}=\frac{p_k^\alpha}{\sum_{i=1}^n p_i^\alpha};\ k=1,\dots,n.
\end{equation}

If $\mcP = \pvec \in \Delta_n$ and $\mcQ = \qvec \in \Delta_m$ the
\textit{direct product}, $\mcP \star \mcQ \in \Delta_{nm}$, is
defined as
\begin{equation}
\mcP \star \mcQ = (p_1q_1, p_1q_2, \dots, p_n q_m).
\end{equation}

The proof of the following theorem can be found in
\cite{Jizba_Arimitsu_04}, \cite{Ilic_et_al_cert_inf}.
\begin{theorem}\rm
\label{Reny 1: theorem}

Let $g:\sR \ra \sR$ be a continuous invertible function and
$\Hv_n: \Delta_n \ra \sR_{>0}$ a continuous function
\begin{equation}
\Hv_{n}(\mcP) = g^{-1} \left( \sum_{k=1}^{n}
\escp{k}(\alpha) g(\cop \cdot \log p_k) \right);  \quad \cop <0, 
\end{equation}
for all $\mcP = (p_1, \dots, p_n) \in \Delta_n$, and let $\Hv_n$
be additive for all $n\in \sN$, i.e.
$\Hv_{nm}(\mcP\star\mcQ)=\Hv_n(\mcP) + \Hv_m(\mcQ)$ for all $\mcP
= (p_1, \dots, p_n) \in \Delta_n$, $\mcQ = (q_1, \dots, q_m) \in
\Delta_m$ $n, m \in \sN$. Then, $g$ is the function from the class
parameterized by $\lambda \in \sR$:
\begin{equation}
\label{Reny 1: theorem: g(x)}
g(x) =
\begin{dcases}
\ \ -cx, &\quad\text{for }\lambda = 0 \\
\frac{2^{-\lambda x}-1}{\gamma}, &\quad\text{for }\lambda \neq 0
\end{dcases}
\quad\quad\Leftrightarrow\quad\quad
g^{-1}(x) =
\begin{dcases}
\ \ -\frac{1}{c}\ x, &\quad\text{for }\lambda = 0 \\
-\frac{1}{\lambda}\log(\gamma x +1),&\quad\text{for }\lambda \neq 0
\end{dcases}
\end{equation}
where $c,\gamma\neq 0$, and the entropy is uniquely determined
with
\begin{eqnarray}
\label{renyi: H with c} \Hv_n(\mcP) =
\begin{dcases}
\cop \cdot
\sum_{k=1}^{n} \escp{k} \log p_k &\quad\text{for }\lambda=0 \\
-\dfrac{1}{\lambda}%
\log \left( \dfrac{\sum_{k=1}^n p_k^{\alpha-\cop
\lambda}}{\sum_{k=1}^n p_k^{\alpha}}\right) &\quad\text{for
}\lambda\neq 0
\end{dcases}
\end{eqnarray}
where $\cop<0$ and $\alpha-\cop \lambda > 0$.
\end{theorem}

%
Nath entropy of $n$-dimensional distribution 
is a function $\Hv_n: \Delta_n \ra \sR_{>0}$ from the family
parameterized by $\alpha,\lambda, \cop \in \sR$:
\begin{align}
\label{Renyi 2: H final c, alpha} \Hv_n(\mcP) =
\begin{dcases}
\ \cop \cdot \sum_{k=1}^{n} p_k  \log p_k ; \quad\quad\quad \ \cop
<0 \quad &\mbox{for} \quad \alpha =1
\\
\ \frac{1}{\lambda} \log \left( \sum_{k=1}^{n} p_k^{\alpha}
\right); \quad \ \frac{1 - \alpha}{\lambda} > 0\ , \ \lambda \neq
0 \quad &\mbox{for} \quad \alpha \neq 1.
\end{dcases}
\end{align}
%
R\'enyi entropy 
is a function $\mc R_n: \Delta_n \ra \sR_{>0}$ from the family
(\ref{Renyi 2: H final c, alpha}) with $\lambda = 1- \alpha$ and
$\cop=-1$.
%
The following theorem characterizes the Nath and R\'enyi
entropies.

\begin{theorem}\rm
\label{Reny 2: theorem} Let the function $\Hv_n: \Delta_n \ra
\sR_{>0}$ satisfies the following axioms, for all $n \in \sN$,
$n>1$:

\begin{description}

\item[NA1:] 
$\Hv_n(\mcU_n)=v(1/n)$, where $v:\sC\ra\sC$ is an analytic
function.
\item[NA2:]
Let $\mcP = (p_1, \dots, p_n) \in \Delta_n$, 
$\mcPQ = (r_{11}, r_{12}, \dots, r_{nm}) \in \Delta_{nm}$, $n, m
\in \sN$ such that $p_i = \sum_{j=1}^n r_{i j}$, and $\mcQ_{ | k}
= (q_{1|k}, \dots, q_{m|k}) \in \Delta_m$, where $q_{i|k} =
r_{ik}/p_k$ and $\alpha \in (0, \infty]$ is some fixed parameter.
Then,
\begin{equation}
\label{Renyi 2: axioms: decomposability}%
\Hv_{nm}(\mcPQ) = \Hv_n(\mcP) + \Hv_m(\mcQ| \mcP),
\quad\text{where}\quad
\Hv_m(\mcQ| \mcP) = f^{-1} \left(\sum_k \escp{k} f(\Hv_m(\mcQ_{ |
k})) \right),
\end{equation}
where $f$ is invertible continuous function.

\end{description}
Then, $\Hv_n$ is the Nath entropy. If, in addition, normalization
axiom $v(1/2)=1$ is satisfied, $\Hv_n$ is the R\'enyi entropy.

\end{theorem}
%
\textbf{Proof.} Let $\mcA=(1/2,1.2)$. By successive use of formula
(\ref{Renyi 2: axioms: decomposability}) we get
\begin{equation}
v\bigg(\frac{1}{2^n}\bigg)={\Hv_{2^n}}(\underbrace{{\mcA} \star
{\mcA} \star \ldots \star
{\mcA}}_{n \text{ times}}) =%
\sum_{k=1}^n \Hv_2(\mcA) =%
n \cdot \Hv_2(\mcA) =%
-\log \frac{1}{2^n} \cdot
{\Hv_2}\bigg(\frac{1}{2},\frac{1}{2}\bigg)=%
\cop \cdot\log \frac{1}{2^n}
\end{equation}
where $\cop=-{\Hv_2}(1/2,1/2)$ is a negative real value, since
${\Hv_2}(1/2,1/2)$ is positive by assumption of the theorem.
Accordingly, the values of functions
$v(z)
$ and $\cop \cdot \log z$ coincide at an infinite number of points
converging to zero, $\{z=1/2^n\}_{n \in \sN}$. Since both $f(z)$
and $\cop \cdot \log z$ are analytic functions, they must be the
same,
\begin{equation}
v(z)=\cop \cdot \log z; \quad \cop<0.
\end{equation}
Let us determine the entropy form for the distribution $\mcP =
(p_1, \dots, p_n) \in \sC^n$ when $p_i$ are rational numbers and the case for irrational numbers follows from the continuity
of entropy. Let  $\mcP = (p_1, \dots, p_n) \in \sC^n$ , $\mcQ_{|k}
= (q_{1|k}, \dots, q_{m_k|k}) \in \sC^{m_k}; k=1,\dots,n$ and
$\mcPQ = (r_{11}, r_{12}, \dots, r_{nm}) \in \sC^{nm}$, for $n, m
\in \sN$, and %
$p_i=m_i/m$; $\quad r_{ij}=1/m$, $q_{j|i}=1/m_i$,
where $m=\sum_{i=1}^n m_i$ and $m_i \in \sN$ for any $i = 1,
\dots, n$ and $j = 1,\dots,m_i$. Then we have
$\Hv_{m}(\mcP\mcQ)=\Hv_{m}(\mcU_{m})=v(1/m) = -\cop \cdot \log m$
and $\Hv_{m_k}(\mcQ_{|k})=\Hv_{m_k}(\mcU_{m_k})=v(1/m_k)= -\cop
\cdot \log m_k$. Since $p_i = \sum_{j=1}^n r_{i j}$, and $q_{i|k}
= r_{ik}/p_k$, we can apply the axiom [\textbf{NA2}] and we get
\begin{equation}
\label{Reny 1: theorem: ra2 for rational} \Hv_n(\mcP) =
-\cop \cdot \log m -%
f^{-1}\left(\sum_{k=1}^n \escp{k}
f\left(-\cop \cdot \log m_k\right)\right) =
-\cop \cdot \log m -f^{-1} \left( \sum_{k=1}^n \escp{k}
f\left(-\cop \cdot \log p_k - \cop \cdot \log m
\right)\right)\, .
\end{equation}
Let us define $f_y(x) = f(-x-y)$, and $f^{-1}_y(z) =
-y-f^{-1}(z)$. If we set $y=\cop\cdot\log m$ the equality (\ref{Reny 1:
theorem: ra2 for rational}) becomes
\begin{equation}
\Hv(\mcP) = f^{-1}_{y} \left( \sum_{k=1}^{n}
\escp{k} f_{y}\left(\cop\cdot\log p_k \right) \right);\quad\quad 
\cop<0. \label{lg1}
\end{equation}
Since $f$ is continuous, both $f_y$ and $f_y^{-1}$ are continuous,
as well as the entropy, and we may extend the result (\ref{lg1})
from rational $p_k$'s to any real valued $p_k$'s defined in [0,1].
Now, if the axiom [\textbf{NA2}] is used with $\mcPQ = \mcP \star
\mcQ$, the conditions from the Theorem \ref{Reny 1: theorem} are
satisfied so that $f_y(x)=-cx$, for $\lambda =0$ and $f_y(x) =
(2^{-\lambda x}-1)/\gamma$, for $\lambda\neq 0$.
%
Accordingly, the entropy is uniquely determined by the class
(\ref{renyi: H with c}).
The relationship between the parameters $\alpha$, $\cop$ and
$\lambda$ is determined by use of the axiom [\textbf{NA2}].
%

For $\lambda=0$, since $f_y(x)=f(z)$, where $z=-x-y$, we get $f(z)=f_y(x)=f_y(-z-y)=cz+cy$. If the equality (\ref{renyi: H with c}) is
substituted in [\textbf{NA2}] with $f(z)=cz+cy$,
we get
\begin{equation}
\sum_{k=1}^{n}\sum_{l=1}^{m} \escr{kl}  
\log{(r_{kl})}=
\sum_{k=1}^{n} \escp{k} %
\log{p_k}\ + \
\sum_{k=1}^{n}\sum_{l=1}^{m} \escp{k} \escqc{l}{k}  %
\log{(q_{l|k})} =
\sum_{k=1}^{n}\sum_{l=1}^{m} \escp{k} \escqc{l}{k}  %
\log{(r_{kl})},
\end{equation}
which can be transformed to
\begin{equation}
\label{renyi: lambda zero condition} \sum_{k=1}^n
\rho_k^{(\alpha)} \sum_{j=1}^m q_{j|k}^\alpha=\sum_{k=1}^n
\escp{k} \sum_{l=1}^m q_{l | k}^\alpha
\quad;\quad
\rho_k^{(\alpha)}=%
\frac{\escp{k} \sum_{l=1}^m \escqc{l}{k} \log r_{kl}}%
{\sum_{i=1}^n \escp{i} \sum_{j=1}^m \escqc{j}{i} \log r_{ij}}.%
\end{equation}
The equality (\ref{renyi: lambda zero condition}) holds for all
distributions and we may consider the case $n=m=2$, and the
distributions $\mcP=(1/2,1/2)$, $\mcQ_{|1}=(1,0)$,
$\mcQ_{|2}=(1/2,1/2)$. If we set
$x_1 = {\sum_{l=1}^2 q_{l|1}^\alpha} = 1$,
$x_2 = {\sum_{l=1}^2 q_{l|2}^\alpha} = 2^{1-\alpha}$
and $u_k=\rho_k^{(\alpha)}$, $v_k=\escp{k}$, the equality
(\ref{renyi: lambda zero condition}) can be transformed as
follows:
\begin{equation}
\label{Reny 1: theorem: ux conditions}
u_1 x_1 + u_2 x_2 = v_1 x_1 + v_2 x_2 \quad\Leftrightarrow\quad%
u_1 x_1 + (1-u_1) x_2 = v_1 x_1 + (1-v_1) x_2 \quad\Leftrightarrow\quad%
(u_1-v_1) x_1 = (u_1-v_1) x_2.
\end{equation}
By using of $u_1= \frac{1}{3} \neq v_1=\frac{1}{2}$, we get
$x_1=x_2$, i.e. $1=2^{1-\alpha}$, which implies $\alpha=1$.
Accordingly, the case $\lambda=0$ from (\ref{renyi: H with c})
reduces to the case $\alpha=1$ from (\ref{Renyi 2: H final c,
alpha}). Positivity of entropy implies that $\cop<0$.
%

If $\lambda\neq0$, and the equality (\ref{renyi: H with c}) is
substituted in [\textbf{NA2}],
we get
\begin{eqnarray}
-\frac{1}{\lambda}%
\log \left( \frac{\sum_{k=1}^n \sum_{l=1}^m r_{kl}^{\alpha-\cop \lambda}}{\sum_{k=1}^n \sum_{l=1}^m r_{kl}^{\alpha}}\right)=%
-\frac{1}{\lambda}
\log \left( \frac{\sum_{k=1}^n p_k^{\alpha-\cop \lambda}}{\sum_{k=1}^n p_k^{\alpha}}\right)+%
f^{-1} \left(\sum_k \escp{k} f\left(-\frac{1}{\lambda}%
\log\left(\frac{\sum_{l=1}^m q_{l|k}^{\alpha-\cop
\lambda}}{\sum_{l=1}^m q_{l|k}^{\alpha}} \right)\right)\right),
\label{rensh1}
\end{eqnarray}
where $f(z) =(2^{\lambda (z+y)} - 1) /\gamma$ (since for $z=-x-y$, we have $f(z) =f_y(-z-y)=(2^{\lambda (z+y)} - 1) /\gamma$) or, equivalently,
\begin{equation}
\label{renyi: lambda not zero condition}
\sum_{k=1}^{n}\sum_{l=1}^{m} \escr{kl}
r_{kl}^{-\cop\lambda}=%
\frac{\sum_{k=1}^{n} \escp{k} \ p_k^{-\cop\lambda}}%
{\sum_{k=1}^n \escp{k} \cdot \dfrac{1}{\sum_{l=1}^m \escqc{l}{k}
q_{l|k}^{-\cop\lambda}}}%
\quad \Leftrightarrow \quad
{\sum_{k=1}^n \sum_{l=1}^m \escr{kl}\cdot%
\dfrac{1}{\sum_{j=1}^m
q_{j|k}^{\beta}}}=%
{\sum_{k=1}^n \sum_{l=1}^m r_{kl}^{(\beta)}\cdot%
\dfrac{1}{\sum_{j=1}^m q_{j|k}^{\beta}}},
\end{equation}
where $\beta=\alpha-\cop\lambda$. Similarly to the case
$\lambda=0$ we note that the equality (\ref{renyi: lambda zero
condition}) holds for all distributions and we consider the case
$n=m=2$, and the distributions $\mcP=(1/2,1/2)$,
$\mcQ_{|1}=(1,0)$, $\mcQ_{|2}=(1/2,1/2)$. If we set $x_1 =
\frac{1}{\sum_{l=1}^2 q_{l|1}^\beta} = 1$, $x_2 =
\frac{1}{\sum_{l=1}^2 q_{l|2}^\beta} = 2^{\beta-1}$ and
$u_k=\sum_{l=1}^m \escr{kl} = \frac{\sum_{l=1}^2
q_{l|k}^\alpha}{\sum_{i=1}^2\sum_{l=1}^2 q_{l|k}^\alpha}$,
$v_k=\sum_{l=1}^m r_{kl}^{(\beta)} = \frac{\sum_{l=1}^2
q_{l|k}^\beta}{\sum_{i=1}^2\sum_{l=1}^2 q_{l|k}^\beta}$, the
equality (\ref{renyi: lambda not zero condition}) can be
transformed into the form (\ref{Reny 1: theorem: ux conditions}).
By using of $u_1=\frac{1}{1+2^{1-\alpha}} \neq
v_1=\frac{1}{1+2^{1-\beta}}$, we get $x_1=x_2$, i.e.
$1=2^{\beta-1}$, which implies $\beta=1$, i.e. $\alpha -
\cop\lambda=1$ with $\cop<0$, and the case $\lambda\neq0$ from
(\ref{renyi: H with c}) reduces to the case  $\alpha\neq1$ from
(\ref{Renyi 2: H final c, alpha}). Positivity of entropy implies
$(1-\alpha)/\lambda > 0$.

Finally, if $v(1/2)=\Hv_2(1/2)=1$, the equation (\ref{Renyi 2: H
final c, alpha}) implies $\cop=-1$, $\lambda=1-\alpha$, which proves
the theorem.

\newcommand{\lama}{\lambda}
\newcommand{\lamq}{\lambda}


\section{Tsallis entropy}
\label{sec: Tsallis}

Havrda-Charv\'at-Tsallis entropy of $n$-dimensional distribution
is a function from the family parameterized by $\cop, \lambda,
\alpha \in \sR$:
\begin{align}
\label{Tsallis: theorem: final form} \mcT(\mcP) =
\begin{dcases}
\ \cop \cdot \sum_{k=1}^{n} p_k  \log \ p_k ; \quad \ \cop <0,
\quad &\mbox{for} \quad a=0
\\
\ \frac{1}{\lama} \cdot \left(\sum_k p_k^\alpha - 1 \right); \quad
\ \alpha\neq1, &\mbox{for} \quad a \neq 0
\end{dcases}
\end{align}
For $\lama=2^{1-\alpha}-1$ the entropy reduces to the
Havrda-Charv\'at entropy \cite{Havrda_Charvat_67}, while in the
case of $\lama=1-\alpha$ it reduces to the Tsallis entropy
\cite{Tsallis_88}.
Let us define
$\op_\lama$-addition,
defined as \cite{Nivanen_Wang__03},
\begin{equation}
\label{SM: op} x \op_\lama y = x + y +\lama x y;\quad a \in \sR,
\end{equation}
for all $x,y \in \sR$. For the case $\lama=0$,
$\op_\lama$-addition reduces to ordinary addition. The pair $(\sR,
\oplus_\lama)$ forms commutative group where the inverse operation
and the $\ominus_\lama$-difference are defined as
\begin{equation}
\ominus_\lama x = \frac{-x}{1+\lama x} \quad\quad\quad x
\ominus_\lama y = \frac{x-y}{1+\lama y}.
\end{equation}
It is easy to see that the structure $(\sR, \op_\lama)$ is a
topological group isomorphic to $(\sR,+)$ with an isomorphism
%
\begin{equation}
\label{SM: Mp}
h(x) = %
\begin{dcases}
\quad x, &\mbox{ for } \lama=0 \\
\frac{ 2^{\lamq \cdot x} - 1}{\lama},
&\mbox{ for } \lama \neq 0
\end{dcases}
\quad\quad\Leftrightarrow\quad\quad
h^{-1}(y) = %
\begin{dcases}
\quad y, &\mbox{ for } \lama = 0 \\
\frac{\log \left(\lama \cdot y + 1 \right)}{\lamq},
&\mbox{ for } \lama \neq 0
\end{dcases}
\end{equation}
so that,
\begin{equation}
\label{tsallis: h is isomorphis}
\isoRtoX(x + y)=\isoRtoX(x)\op_\lama\isoRtoX(y)%
\quad\quad\Leftrightarrow\quad\quad \isoXtoR(x \op_a
y)=\isoXtoR(x)+\isoXtoR(y).
\end{equation}
The following theorem characterizes the Havrda-Charv\'at-Tsallis
entropy.

\begin{theorem}\rm
\label{Tsallis 2: theorem}

Let the function $\mcT_n: \Delta_n \ra \sR_{>0}$ satisfies the
following axioms, for all $n \in \sN$, $n>1$:

\begin{description}

\item[TA1:] 
$\mcT_n(\mcU_n)=t(1/n)$, where $t:\sC\ra\sC$ is an analytic
function.
\item[TA2:]

Let $\mcP = (p_1, \dots, p_n) \in \Delta_n$, 
$\mcPQ = (r_{11}, r_{12}, \dots, r_{nm}) \in \Delta_{nm}$, $n, m
\in \sN$ such that $p_i = \sum_{j=1}^n r_{i j}$, and $\mcQ_{ | k}
= (q_{1|k}, \dots, q_{m|k}) \in \Delta_m$, where $q_{i|k} =
r_{ik}/p_k$. Then
\begin{equation}
\label{Tsallis 2: axioms: decomposability}%
\mcT_{nm}(\mcPQ) = \mcT_n(\mcP) \op_\lama \mcT_m(\mcQ| \mcP),
\quad\text{where}\quad
\mcT_m(\mcQ| \mcP) =\sum_k \escp{k} \mcT_m(\mcQ_{ | k})
\end{equation}
\end{description}
Then, $\mcT_n$ is the Havrda-Charv\'at-Tsallis entropy.
\end{theorem}

\newcommand{\sgn}{\mbox{sgn}}
\textbf{Proof.} If $\lama=0$, the Theorem \ref{Tsallis 2: theorem}
reduces to the Theorem \ref{Shannon: theorem}, so we prove the
theorem for $\lama\neq 0$. Similarly as in the proof of the
Theorem \ref{Reny 2: theorem}, we set $\mcA=(1/2,1.2)$. By
successive usage of formula (\ref{Tsallis 2: axioms:
decomposability}), we get
\begin{equation}
t\bigg(\frac{1}{2^n}\bigg)={\mcT_{2^n}}(\underbrace{{\mcA} \star
{\mcA} \star \ldots \star
{\mcA}}_{n \text{ times}}) =%
\bigoplus_{k=1}^n \mcT_2(\mcA) =%
\isoRtoX\left(\sum_{k=1}^n \mcT_2(\mcA) \right)=
\isoRtoX\Big(n \cdot \mcT_2(\mcA)\Big) = 
\isoRtoX\left(\cop\cdot\log \frac{1}{2^n}\right),
\end{equation}
where $\cop=-{\mcT_2}(1/2,1/2)$ is a negative real value, since
${\mcT_2}(1/2,1/2)$ is positive, by assumption of the theorem and
$\sgn(h(x))=\sgn(x)$. Accordingly, the values of functions
$t(z)
$ and $\isoRtoX(\cop \cdot \log z)$ coincide at an infinite number of
points converging to zero, $\{z=1/2^n\}_{n \in \sN}$. Since both
$t(z)$ and $h(\cop \cdot \log z)$ are analytic functions
\footnote{Note that in this context, $h$ is an analytic
continuation of the real function defined by the formula (\ref{SM:
Mp}).}, they must be the same,
\begin{equation}
t(z)=\isoRtoX(\cop \cdot \log
z)=\frac{1}{\lama}\left(z^{\cop\lamq}-1\right); \quad \cop<0.
\end{equation}

Similarly as  in the proof of the theorem (\ref{Reny 2: theorem}),
we determine the entropy form for the distribution $\mcP = (p_1,
\dots, p_n) \in \sC^n$ when $p_i$ are rational numbers and the
case for irrational numbers follows from the continuity of the
entropy. Let $\mcP = (p_1, \dots, p_n) \in \sC^n$ , $\mcQ_{|k} =
(q_{1|k}, \dots, q_{m_k|k}) \in \sC^{m_k}; k=1,\dots,n$, and
$\mcPQ = (r_{11}, r_{12}, \dots, r_{nm}) \in \sC^{nm}$, for $n, m
\in \sN$, where $p_i=m_i/m$; $r_{ij}=1/m$, and $q_{j|i}=1/m_i$;
where $m=\sum_{i=1}^n m_i$ and $m_i \in \sN$, for any $i = 1,
\dots, n$ and $j = 1,\dots,m_i$. Then, $p_i = \sum_{j=1}^n r_{i
j}$, and $q_{i|k} = r_{ik}/p_k$ and we can apply the axiom
[\textbf{TA2}], which gives

\begin{equation}
\label{Tsallis 2: theorem: T with c}
\mcT(\mcP) = t\left(\frac{1}{m}\right)  \ominus_\lama \sum_{k=1}^n p_k^\alpha\ t\left(\frac{1}{m_k}\right)  =%
\frac{t\left(\frac{1}{m}\right)  - \sum_{k=1}^n \escp{k}\
t\left(\frac{1}{m_k}\right) }{1 + \lama\cdot
\sum_{k=1}^n \escp{k}\ t\left(\frac{1}{m_k}\right) }=%
\frac{1}{\lama}\cdot\left(\frac{\sum_{k=1}^n p_k^\alpha}{
\sum_{k=1}^n p_k^{\alpha-\cop \lamq}} - 1 \right).
\end{equation}

The relationship between $\cop$ and $\lamq$ is determined by use
of the axiom [\textbf{TA2}]. If we apply the map $h$ on both sides
of the equality (\ref{Tsallis 2: axioms: decomposability}), using
the equality (\ref{tsallis: h is isomorphis}), by which
$\isoXtoR(x \op_a y)=\isoXtoR(x)+\isoXtoR(y)$, we get
$\isoXtoR(\mcT(\mcPQ)) = \isoXtoR(\mcT(\mcP))+ \isoXtoR(\sum_k
\escp{k} \mcT(\mcQ_{ | k}))$. If we now use the equalities
(\ref{SM: Mp}) for the mappings $h$ and $h^{-1}$ and the entropy
form (\ref{Tsallis 2: theorem: T with c}), we get
\begin{equation}
\label{sec: tsallis: q c conditions}
\frac{1}{\lamq}\cdot\log\left(\frac{\sum_{k=1}^n r_{kl}^{\alpha}}{
\sum_{k=1}^n r_{kl}^{\alpha-\cop\lamq}} \right)=
\frac{1}{\lamq}\cdot\log \left(\frac{\sum_{k=1}^n p_k^\alpha}{
\sum_{k=1}^n p_k^{\alpha-\cop\lamq}}\right)
+h^{-1}\left(\sum_{k=1}^n \escp{k}\ h\left(\frac{1}{\lamq}\cdot
\log \left(\frac{\sum_{k=1}^n q_{l|k}^\alpha}{ \sum_{k=1}^n
q_{l|k}^{\alpha-\cop\lamq}} \right)\right)\right).
\end{equation}

The function $h(z)=(2^{\lamq \cdot z} - 1)/\lamq$ is the linear
function of $f(z) =(2^{\lamq (z+y)} - 1) /\gamma$. It is a
well-known fact from the mean theory that, if $h$ is a linear
function of $f$, they generate the same quasi-linear mean
\cite{Hardy_et_al_34}, and the function $h$ in the equality
(\ref{sec: tsallis: q c conditions}) can be substituted with $f$.
Accordingly,
the equation (\ref{sec: tsallis: q c conditions}) can be rewritten
in the form of the equation (\ref{rensh1}). 
As shown in the proof of the Theorem \ref{Reny 2: theorem}, the
equation (\ref{rensh1}) is satisfied iff $\alpha-\cop\lambda=1$,
and the equation (\ref{Tsallis 2: theorem: T with c}) reduces to
the equation (\ref{Tsallis: theorem: final form}), which proves
the theorem.

%

\bibliographystyle{ieeetr}
\bibliography{reference}

\end{document}